\documentclass[aps,prl,reprint,superscriptaddress,longbibliography, footinbib]{revtex4-1}
\usepackage[utf8]{inputenc}
\usepackage{graphicx}
\usepackage{dcolumn}
\usepackage{bm}
\usepackage{amsmath}
\usepackage{color}
\usepackage[usenames,dvipsnames,svgnames,table]{xcolor}
\usepackage{textcomp}
\newcommand{\textapprox }{\raisebox{0.5ex}{\texttildelow}}
\usepackage[colorlinks=true,
linkcolor=black,
urlcolor=black,
citecolor=black]{hyperref}
\usepackage{siunitx}

\newcommand{\ket}[1]{\ensuremath{\left|{#1}\right\rangle}}
\newcommand{\Rb}[1]{\ensuremath{^{#1}}\textrm{Rb}}

\graphicspath{{figures/}}
\begin{document}
	
	\title{Continuous quantum light from a dark atom}
	
	\author{Karl Nicolas Tolazzi}
	\email{nicolas.tolazzi@mpq.mpg.de}
	\author{Bo Wang}
	\author{Christopher Ianzano}
	\author{Jonas Neumeier}
	\affiliation{Max-Planck-Institut f\"ur Quantenoptik, Hans-Kopfermann-Str.\ 1, D-85748 Garching, Germany}
	\author{Celso Jorge Villas-Boas}
	\affiliation{Departamento de F\'isica, Universidade Federal de S\~ao Carlos, P.O. Box 676, 13565-905, S\~ao Carlos, Brazil}
	\author{Gerhard Rempe}
	\affiliation{Max-Planck-Institut f\"ur Quantenoptik, Hans-Kopfermann-Str.\ 1, D-85748 Garching, Germany}
	
	\date{\today}
	
	\maketitle

	\subsection{Abstract}
	Cycling processes are important in many areas of physics ranging from lasers to topological insulators, often offering surprising insights into dynamical and structural aspects of the respective system. Here we report on a quantum-nonlinear wave-mixing experiment where resonant lasers and an optical cavity define a closed cycle between several ground and excited states of a single atom. We show that, for strong atom-cavity coupling and steady-state driving, the entanglement between the atomic states and intracavity photon number suppresses the excited-state population via quantum interference, effectively reducing the cycle to the atomic ground states. The system dynamics then result from transitions within a harmonic ladder of entangled dark states, one for each cavity photon number, and a quantum Zeno blockade that generates antibunching in the photons emitted from the cavity. The reduced cycle suppresses unwanted optical pumping into atomic states outside the cycle, thereby enhancing the number of emitted photons.

	\subsection{Introduction}
	When two light fields resonantly drive transitions from two different atomic ground states to a common excited state, forming a three-level lambda-type system, the quantum-mechanical transition amplitudes to that state can interfere destructively. This inhibits excitation and produces a superposition between the two ground states, called a dark state \cite{Alzetta1976}. Dark states give rise to electromagnetically induced transparency \cite{Harris1990} and slow light \cite{Kasapi1995}, appear in lasing without inversion \cite{Mompart2000}, are employed for producing \cite{Kuhn2002} and storing \cite{Specht2011} single photons in quantum networks \cite{Ritter2012,Reiserer2015}, and are widely used in physics and chemistry in the form of a stimulated Raman adiabatic passage \cite{Vitanov2017}.
	
	The main advantage of such dark states is that they protect the system from decoherence related to the excited state. They are thus ideal for continuous experiments with atomic cycling currents involving the excited state, as the removal of the decay channel allows the cycle to run for a longer period. Unfortunately, attempting to close the lambda system and generating a cycling current with a third field that directly couples the two ground states results in the destruction of the dark state, except in the restricting case when both driving strengths of the lambda subsystem are equal \cite{Pope2019}. In general, the breakdown of the dark state in such a system brings population to the excited atomic state. Thus, decoherence is reintroduced via atomic decay and additional loss channels are opened via depumping to, e.g., states outside the cycle.
	
	As we report here, this breakdown can be mitigated by replacing one of the coherent driving fields of the closed cycle by an optical cavity strongly coupled to the corresponding atomic transition. In this case the entanglement of the atomic ground states with the photon number in the cavity effectively preserves the destructive interference of atomic excitation amplitudes, and thus, the dark states. The possibility to have any number of photons in the cavity furthermore produces an infinite harmonic ladder of dark states that can be used to produce either quantum or coherent light \cite{Villas-Boas2019}. 
	Replacing a laser with a cavity has the further advantage that it introduces in the otherwise decoherence-free subspace in which the system operates, a well-defined dissipation channel through which the system dynamics can be observed in real time.

	\subsection{Results and Discussion}
	Our system consists of a single \Rb{87} atom strongly coupled to a high-finesse optical Fabry-P\'erot resonator. Our parameters are $(\kappa,\gamma,g)/2\pi=(1.5,3.0,10.2)\,\text{MHz}$, with $\kappa$ the cavity-field decay rate, $\gamma$ the atomic polarization decay rate, and $g$ the atom-cavity coupling constant. This $g$ is the maximum value and varies slightly between atoms at different trapping positions within the cavity. The condition $(\kappa,\gamma)\ll g$ puts the system well into the strong-coupling regime of cavity quantum electrodynamics (CQED).
	
\begin{figure}[t]
		\includegraphics[width=8.6cm]{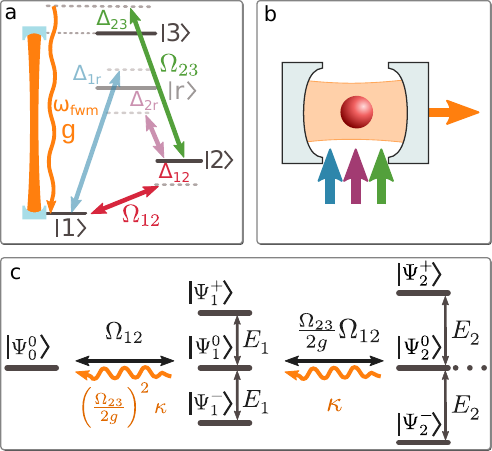}
		\caption{\label{fig_lvl_scheme}System description. (a) shows the atomic level scheme with ground states \ket{1} and \ket{2}, the intermediate virtual state \ket{r} and excited state \ket{3}. The atom is strongly coupled to the cavity (orange) which is resonant to the transition from \ket{1} to \ket{3}. The system is driven by three lasers with  detunings $\Delta_{1r}, \Delta_{2r}, \Delta_{23}$ to their respective excited states. A new field with frequency $\omega_{\text{fwm}}$ is generated. (b) shows a sketch of the experimental setup including the atom, cavity and light fields that drive the atom. (c) shows the effective level scheme that forms for $\Omega_{12} \ll (\Omega_{23},g)$, where the dark state transitions are depicted with their driving strength and their decay rates in the limit of $\Omega_{23} \ll g$.}
	\end{figure}

	Fig. \ref{fig_lvl_scheme}(a) shows the relevant level scheme of the atom together with the driven transitions. Two ground states $\ket{1}=\ket{5\text{S}_{1/2},\text{F}=1,\text{m}_{\text{F}}=1}$ and $\ket{2}=\ket{5\text{S}_{1/2},\text{F}=2,\text{m}_{\text{F}}=2}$ are connected via a Raman transition with intermediate state $\ket{r}$ which is a virtual level close to the $D_{1}$ line of Rubidium (with detuning $\textapprox \SI{4.5}{G\hertz}$). The cavity is resonant with the transition $\ket{1}\leftrightarrow\ket{3}=\ket{5\text{P}_{3/2},\text{F}=2,\text{m}_{\text{F}}=2}$. In the experiment, the transitions $\ket{1}\leftrightarrow\ket{r}$, $\ket{2}\leftrightarrow\ket{r}$ and $\ket{2}\leftrightarrow\ket{3}$ are addressed by lasers with frequencies $\omega_{1r}$, $\omega_{2r}$, and $\omega_{23}$ respectively, all of them illuminating the atom. Together with the transition driven by the vacuum field of the cavity, this forms a closed cycle where new light with frequency $\omega_{\text{fwm}}$ is continuously generated and emitted from the cavity. The scheme can be seen as a double-lambda or butterfly system (a typical system capable of four-wave mixing processes \cite{Merriam2000,Dubin2010}) where the two lambda subsystems share ground states. As the Raman transition is far off-resonant we can assume a simplified level scheme where the transition between levels $\ket{1}$ and $\ket{2}$ is resonantly driven with Rabi frequency $\Omega_{12}$ and the state $\ket{r}$ is eliminated.

	For all fields resonant and $\Omega_{12}=0$, the eigenstates of the system form a ladder of triplets (except for the lowest state) (see Fig. \ref{fig_lvl_scheme}(c)), as in cavity electromagnetically induced transparency \cite{Mucke2010,Kampschulte2010a,Albert2011,Souza2013}. These states then read
	\begin{align}
		\ket{\Psi_0^0}&= \ket{1,0}
	\end{align}
	for zero photons and
	\begin{align}
		\ket{\Psi_n^0}&\propto \frac{\Omega_{23}}{2}\ket{1,n}-g\sqrt{n}\ket{2,n-1},\\
		\ket{\Psi_n^\pm}&\propto g\sqrt{n}\ket{1,n}+\frac{\Omega_{23}}{2}\ket{2,n-1}\pm E_n\ket{3,n-1},
	\end{align}
	with $E_n=\sqrt{ng^2 + \Omega_{23}^2/4}$ for $n$ photons. Here the bare states are of the form \ket{\text{atomic state},\text{photon number}}, and $\Omega_{23}$ is the Rabi frequency of the transition $\ket{2}\leftrightarrow\ket{3}$. The states $\ket{\Psi_n^\pm}$ are superposition states with both atomic and cavity excitation and have an energy splitting of $E_n$. The states $\ket{\Psi_n^0}$ are dark states with no contribution from the excited atomic state, $\ket{3}$, exhibiting atom-cavity entanglement for $n>0$. For $\Omega_{12}\ll(g,\Omega_{23},\kappa)$, as in our experiment, the coherent Raman-laser driving can be treated as a perturbation. 
	
\begin{figure}[t]
		\includegraphics[width=8.6cm]{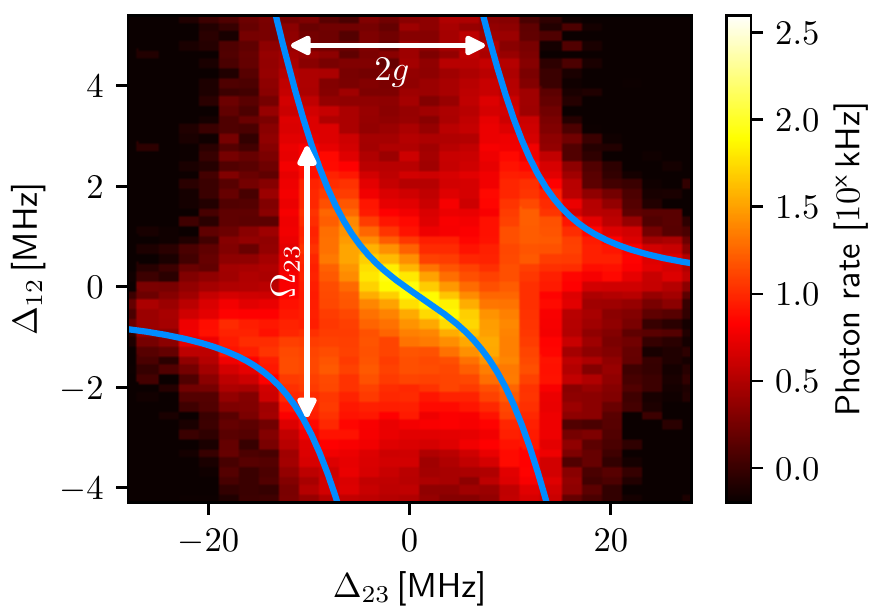}
		\caption{\label{fig_eigenenergies}System spectroscopy. The emitted rate of photons from the cavity versus $\Delta_{12}$ and $\Delta_{23}$ in logarithmic scale. The blue lines in the figure are the theoretically calculated energies of the one-photon states for the ideal system. Also marked are the splittings between these eigenenergies, which are $2g$ in horizontal
			and $\Omega_{23}$ in the vertical
			direction. The experimental Rabi frequencies are  $\Omega_{23}/2\pi=\SI{4.0}{\mega\hertz}$ and $\Omega_{12}/2\pi=\SI{0.4}{\mega\hertz}$ with a coupling constant  $g/2\pi=\SI{10.2}{\mega\hertz}$.}
	\end{figure}

	The eigenenergies of the system can be measured by driving transitions from the ground state to these dressed eigenstates. Only when the driving fields resonantly excite an eigenstate of the first (or higher) manifold, the system exhibits a nonzero cavity photon number and photons leave the cavity. Fig. \ref{fig_eigenenergies} shows the photon emission rate from the cavity when performing a two-dimensional scan of the detuning $\Delta_{12}$ against $\Delta_{23}$. This scan reveals the complex energy landscape in the first manifold that shows two avoided crossings. The highest photon generation rate is reached on resonance, $\Delta_{12}=\Delta_{23}=0$, with a rate of the order of \SI{100}{\kilo\hertz}.

	\begin{figure}[t]
		\includegraphics[width=8.6cm]{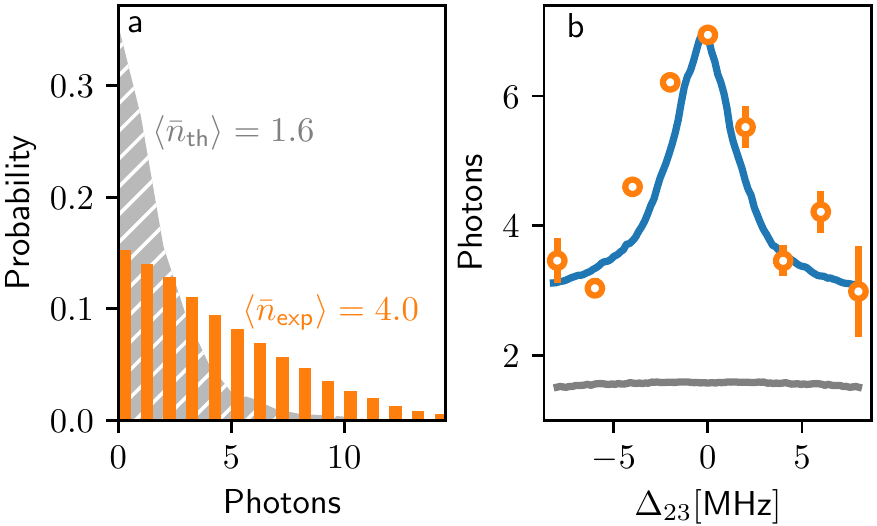}
		\caption{\label{fig_excitationless}Photon generation from a dark state. (a) shows the photon-number probability distribution (orange) measured for zero detuning ($\Delta_{23}=\Delta_{12}=0)$ and a \SI{60}{\micro\second} long observation interval. As a comparison, a quantum simulation was performed to find the expected photon-number distribution (gray curve in background) from a free-space atom interacting with the same driving fields. (b) shows the experimentally measured number of photons (orange), extrapolated to infinite measurement time as a function of $\Delta_{23}$. The curve (blue) is the number of photons from the quantum Monte-Carlo simulation of the experiment. It serves for qualitative comparison, is scaled in the vertical direction, and an offset is added to compensate for experimental imperfections. The gray curve shows the full quantum Monte-Carlo simulation for a free-space atom.}
	\end{figure}

	To demonstrate that these photons stem from transitions between dark states, and therefore do not suffer from atomic decay, the figure of merit is the ratio of photons produced in the cavity per residual atomic excitation. This number is proportional to the total number of photons the system can generate before it decays into an uncoupled state (see Supplementary Material). Such a channel out of the system is only possible via spontaneous decay from the excited state. 
	To that end, Fig. \ref{fig_excitationless}(a) shows in orange the measured photon-number probability distribution for a measurement time of \SI{60}{\micro\second}. In that interval we measure on average 4.0 photons (out of presumably 16 produced photons, due to the limited total detection efficiency of 26\%). The gray background curve in Fig. \ref{fig_excitationless}(a) shows the photon number from a quantum simulation of an identical level scheme without a cavity (and therefore no strong coupling). This simulation assumes infinite measurement time and a detector capable of detecting photons in a $4\pi$ solid angle around the atom (to account for the loss of directionality due to the lack of a cavity), with the same total quantum efficiency as in the experiment. The total expected number of detected photons before this system is pumped into an uncoupled state is then 1.6, which is a factor of 2.5 smaller than our measured result of 4.0 photons. %
	
	Extending this result, Fig. \ref{fig_excitationless}(b) plots the average number of measured photons as a function of $\Delta_{23}$, extrapolated for infinite measurement times (see Supplementary Material). In the zero-detuning case we find 6.9(1) photons. Comparing this to the cavity-free result of 1.6 photons, our strongly coupled system shows a 4.3-fold increase in the number of photons generated in the more realistic comparison between infinite measurement times. Additionally, a clear peak is seen in the data, demonstrating that the dark state transition occurs exclusively at zero detuning. This contrasts the result of the quantum simulation without a cavity (gray curve in Fig. \ref{fig_excitationless}(b)), which is essentially a flat line at around 1.6 photons.

\begin{figure}[t]
		\includegraphics[width=8.6cm]{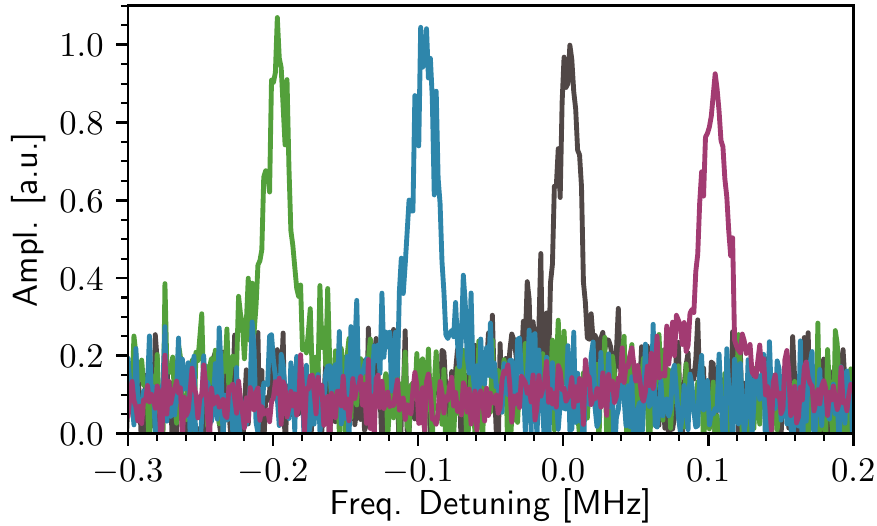}
		\caption{\label{fig_shift}Quantum nonlinear optics. This shows the spectrum of the cavity output field as measured via a heterodyne setup for different input field detunings. The starting point is the black curve with  $\Delta_{1r}=\Delta_{2r}=\Delta_{23}=0$. The other curves are detuned by $\Delta_{1r}=\SI{100}{\kilo\hertz}$ for the blue, $\Delta_{2r}=\SI{100}{\kilo\hertz}$ for the violet, $\Delta_{23}=\SI{200}{\kilo\hertz}$ for the green. The measured amplitude is normalized to the largest amplitude.}
	\end{figure}

	We characterize the generated field with a heterodyne detection setup that is used as a spectrum analyzer as described in the methods section. Different spectra of the newly generated light field are shown in  Fig. \ref{fig_shift}.

	The spectral width of the light field is around \SI{20}{kHz} for a measurement time of \SI{55}{\micro\second}.  This small width can be explained by the fact that all drive lasers, including the local oscillator which is used to measure the spectrum, are phase locked to a common frequency comb that serves as a reference. This measurement thus shows that the newly generated field is phase coherent to all input lasers with a remarkable coherence length. To investigate how the new light field behaves under change of the input light fields, we tune the frequency of these sequentially as seen in the green, blue and violet plot in Fig. \ref{fig_shift}. When tuning the frequency of the input field, the output field changes its frequency according to $\omega_{\text{fwm}}=\omega_{1r}+\omega_{23}-\omega_{2r}$,
	reflecting energy conservation. This shows that the system indeed mediates a nonlinear four-wave-mixing process with one single atom.
	
	To investigate the ladder of dark states in greater detail, it is helpful to derive the interaction Hamiltonian of the system. For all fields being resonant ($\Delta_{12}=\Delta_{23}=0$), and in the regime $\Omega_{23} \ll g$ the interaction Hamiltonian simplifies to (with $\hbar=1$) \cite{Villas-Boas2019}
	\begin{align}\label{hamiltonian_interaction}
		H_{\text{int}} \simeq - \frac{\Omega_{12}}{2}|\Psi_{1}^0\rangle \langle\Psi_{0}^0|-\sum_{n=1}^{\infty}\frac{\Omega_{12}\Omega_{23}}{4 g\sqrt{n}}|\Psi_{n+1}^0\rangle \langle\Psi_{n}^0|+h.c.
	\end{align}
	For higher $\Omega_{23}$ this expression becomes lengthy and can be found in the Supplementary Material. 
	
	The Hamiltonian in Eq. \ref{hamiltonian_interaction} describes the generation of photons in the cavity mode via transitions between different dark states that avoid exciting the atom. It also shows that although transitions between each pair of subsequent dark states in the ladder are possible, the transition strengths and the decay rates of the dark states (as Fig. \ref{fig_lvl_scheme}(c) shows) change in a highly nonlinear fashion with the excitation rung. In fact, the decrease in driving strength to, along with the increase in decay rate from, the higher-lying dark states, results in a restriction of the Hilbert space to lower photon number states, leading to a quantum Zeno effect \cite{Misra1977}. As a result, the system is restricted to the first two dark states, where it behaves like a two-level atom and higher rungs are blocked.

	\begin{figure}[t]
		\includegraphics[width=8.6cm]{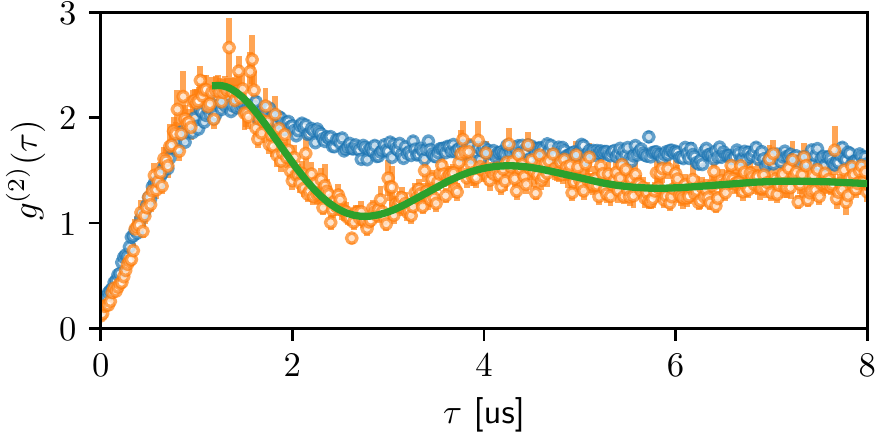}
		\caption{\label{fig_correlation}Photon correlation function. Two-photon correlation function measured with two single-photon detectors in Hanbury Brown and Twiss configuration with a coincidence window of \SI{20}{\nano\second}. Shown are correlation functions for $\Delta_{12}=\Delta_{23}=0$ and driving strengths $\Omega_{23}/2\pi=\SI{2.0}{\mega\hertz}$ (orange dots) and $\Omega_{23}/2\pi=\SI{3.7}{\mega\hertz}$ (blue dots). On top of the orange dots is a sinusoidal fit with exponentially decaying amplitude (green line).}
	\end{figure}

	Insights into the dynamics within the ladder of dark states can be obtained by investigating the photon statistics via the standard correlation function $g^{(2)}(\tau)$. Measured data for two different values of $\Omega_{23}$ are shown in Fig. \ref{fig_correlation}. They display strong photon antibunching and pronounced sub-Poissonian statistics. This single-photon characteristic shows that the system indeed exhibits a Zeno blockade as higher dark state rungs are strongly suppressed. Also visible is a damped oscillation with frequency of $\Omega_{12}/2\pi$, due to the Rabi oscillation between the lowest dark states $\ket{\Psi_0^0}$ and $\ket{\Psi_1^0}$. The green curve is a fit of a damped oscillation to the data that determines this frequency to \SI{327(14)}{\kilo\hertz} which is in decent agreement with the calculated frequency of \SI{400}{\kilo\hertz}, estimated from the laser powers. 
	
	The two curves in Fig. \ref{fig_correlation} feature the same oscillation frequency and rise time for different $\Omega_{23}$, but the decay constant of the oscillation changes, showing that the lifetime of the dark state and therefore the coherence time depends on $\Omega_{23}$ (see Fig. \ref{fig_lvl_scheme}(c) and the Supplementary Material for the dark state decay rates). Note that the dephasing rate for the Rabi oscillation differs from the linewidth in Fig. \ref{fig_correlation}. This is not a contradiction as first-order coherence does not imply second-order coherence \cite{Hoffges1997}, and indicates that the individual photons are delocalized in a continuous and coherent wave.

	\begin{figure}[t]
		\includegraphics[width=8.6cm]{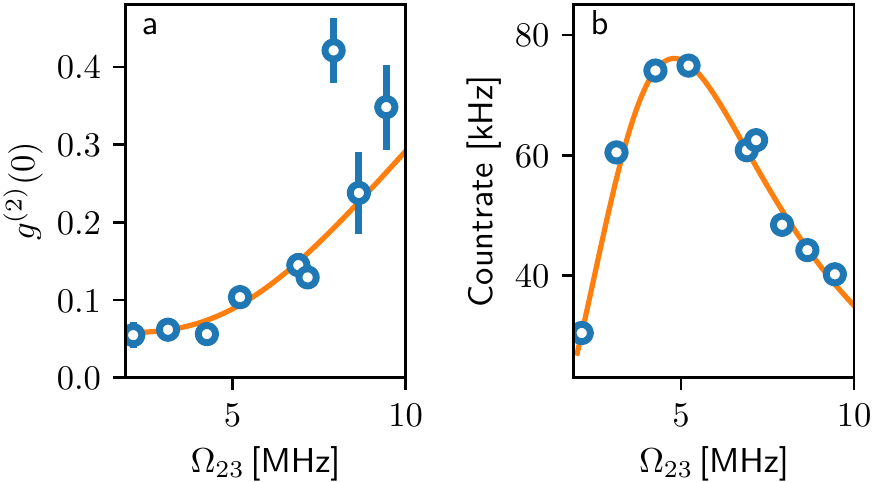}
		\caption{\label{fig_zeno}Zeno blockade. (a) shows the equal-time correlation value for different values of $\Omega_{23}$ for $\Omega_{12}/2\pi=\SI{300}{\kHz}$. (b) shows the measured photon rate of the cavity output for the same measurement. The orange lines show a quantum simulation of the simplified system with a position averaged coupling constant of $g/2\pi=\SI{9.2}{\mega\hertz}$ and a (slight) amplitude correction of the photon production rate to accommodate the data. The error bars are statistical.}
	\end{figure}

	Going beyond the effective two-level system is possible as both the decay rates and the effective driving strength are tunable with $\Omega_{23}$. A higher value of $\Omega_{23}$ allows one to climb the harmonic dark-state ladder and operate the system in a regime where it exhibits a more linear behaviour. In this context, Fig. \ref{fig_zeno}(a) displays the equal-time photon correlation for increasing $\Omega_{23}$ and a fixed $\Omega_{12}$. It shows that $g^{(2)}(0)$ increases with $\Omega_{23}$: the system deviates from a two-level system and approaches the photon statistics of a coherent field when the Zeno blockade is removed and higher-lying dark states become populated. Counterintuitively, a stronger control field does not always increase and in fact even decreases the number of emitted photons, as displayed in Fig. \ref{fig_zeno}(b). This behaviour is predicted by the full analytic formula for the effective driving strength given in the Supplementary Material. It proves that the change of the photon statistics does not simply reflect a buildup of photons in the cavity, but instead is an effect of turning the system into a linear one.

	The closed cycling scheme presented here is generic and could be implemented in all strongly coupled cavity systems to suppress emitter excitation. This is particularly relevant, for instance, in the fluorescence observation of molecules, where excitation leads to rapid depumping to uncoupled rovibrational states. In addition to creating more photons, our method has the additional advantage that it provides a background-free signal as the newly generated field is of completely different frequency than the input drivings. Further prospects are opened up by increasing the driving strength $\Omega_{12}$ between the two ground states. This can result, e.g., in the generation of Schr\"odinger-cat states of light \cite{Villas-Boas2019}. Our scheme is also interesting for the investigation of cycles that simulate Hamiltonians from many-body physics under continuous observation \cite{Laflamme2017}. Finally, closed cycles driven by light fields at the single-photon level could offer intriguing possibilities for the realization of quantum heat engines where cavity fields could act as non-classical heat baths along the thermodynamic cycle \cite{Harbola2012}.
	
	\subsection{Methods}
	The aforementioned experiments are performed with a single $^{87}$Rb atom trapped inside a high finesse resonator using a three-dimensional optical lattice. This leads to typical trapping times of around \SI{6}{\second} during which measurements are taken at a rate of about \SI{1}{\kilo\hertz}. A detailed description of the experimental trapping setup and information about the used resonator are given in \cite{HamsenThesis}.
	A typical experimental sequence consists of a cooling interval followed by optical pumping to a specific Zeeman substate (e.g. $\ket{1}$ as defined in the results section) and the subsequent experiment. 
	
	We investigate the newly generated field in the experiment via two single photon detectors in a Hanbury Brown and Twiss configuration for intensity and correlation measurements or in the spectral domain by heterodyne detection. The heterodyne detection method  employs two photodiodes and a strong local oscillator beam (LO) at frequency $\omega_{\text{LO}}$ that is superimposed with the light field to be measured (called probe; at frequency $\omega_{\text{p}}$), producing an electronic beat signature at frequency $\omega_{\text{b}}=|\omega_{\text{LO}}-\omega_{\text{p}}|$. Using this scheme, it is possible to detect light with extremely low power by optical amplification of the probe caused by the LO. In contrast to a homodyne scheme, where $\omega_{LO}=\omega_{p}$, LO and probe have separate frequencies, thus circumventing detector noise which predominantly occurs at low frequencies. Moreover, using balanced photodiodes we cancel classical intensity fluctuations, making the setup more sensitive. The electronic demodulation process after the output of the balanced photodiodes generates two signals, each of which is proportional to one of the field quadratures of the probe. The demodulation is based on mixing the sinusoidal beat between probe and LO with an electronic local oscillator wave (eLO) at frequency $\omega_{\text{eLO}}$. In addition, we implement a low pass filter with cutoff frequency $\omega_{\text{LP}}$ to remove high frequency noise after the downmixing process. With the two quadrature signals, we can deduce the amplitude and phase of the probe beam. The heterodyne setup is here predominantly used as a spectrum analyzer. When recording a spectrum, we fix $\omega_{\text{eLO}}$ and vary $\omega_{\text{LO}}$ with an AOM. A signal can only be observed if the following condition is satisfied: $|\omega_{\text{eLO}}-\omega_{\text{b}}|<\omega_{\text{LP}}$. This represents a scanning window which produces a non-zero output signal if the probe frequency $\omega_p$ lies within the window.

	\subsection{Acknowledgements}
	B.W. was supported by Elitenetzwerk Bayern (ENB) through the doctoral program ExQM. K.N.T. was supported by the Deutsche Forschungsgemeinschaft (DFG) under Germany’s Excellence Strategy – EXC-2111 – 390814868. C.J.V.-B. thanks the support by the Sao Paulo Research Foundation (FAPESP) Grants No. 2018/22402-7 and No. 2019/11999-5, and National Council for Scientific and Technological Development (CNPq) Grants No. 307077/2018-7 and No. 465469/2014-0
	
	\subsection{Author contributions}
	All authors contributed to the theoretical understanding, the analysis of the measurement, and the writing of the manuscript. 
	
	\subsection{Competing interests}
	The authors declare no competing financial interests.
	
	\subsection{Data availability}
	Data is available from the authors upon request.

	\bibliographystyle{naturemag}

\begin{thebibliography}{10}
		\expandafter\ifx\csname url\endcsname\relax
		\def\url#1{\texttt{#1}}\fi
		\expandafter\ifx\csname urlprefix\endcsname\relax\def\urlprefix{URL }\fi
		\providecommand{\bibinfo}[2]{#2}
		\providecommand{\eprint}[2][]{\url{#2}}
		
		\bibitem{Alzetta1976}
		\bibinfo{author}{Alzetta, G.}, \bibinfo{author}{Gozzini, A.},
		\bibinfo{author}{Moi, L.} \& \bibinfo{author}{Orriols, G.}
		\newblock \bibinfo{title}{An experimental method for the observation of r.f.
			transitions and laser beat resonances in oriented {Na} vapour}.
		\newblock
		\href{http://dx.doi.org/10.1007/BF02749417}{\emph{\bibinfo{journal}{Il Nuovo
					Cimento B}} \textbf{\bibinfo{volume}{36}}, \bibinfo{pages}{5--20}}
		(\bibinfo{year}{1976}).
		
		\bibitem{Harris1990}
		\bibinfo{author}{Harris, S.~E.}, \bibinfo{author}{Field, J.~E.} \&
		\bibinfo{author}{Imamo{\u g}lu, A.}
		\newblock \bibinfo{title}{Nonlinear optical processes using electromagnetically
			induced transparency}.
		\newblock
		\href{http://dx.doi.org/10.1103/PhysRevLett.64.1107}{\emph{\bibinfo{journal}{Physical
					Review Letters}} \textbf{\bibinfo{volume}{64}}, \bibinfo{pages}{1107--1110}}
		(\bibinfo{year}{1990}).
		
		\bibitem{Kasapi1995}
		\bibinfo{author}{Kasapi, A.}, \bibinfo{author}{Jain, M.}, \bibinfo{author}{Yin,
			G.~Y.} \& \bibinfo{author}{Harris, S.~E.}
		\newblock \bibinfo{title}{Electromagnetically {Induced} {Transparency}:
			{Propagation} {Dynamics}}.
		\newblock
		\href{http://dx.doi.org/10.1103/PhysRevLett.74.2447}{\emph{\bibinfo{journal}{Physical
					Review Letters}} \textbf{\bibinfo{volume}{74}}, \bibinfo{pages}{2447--2450}}
		(\bibinfo{year}{1995}).
		
		\bibitem{Mompart2000}
		\bibinfo{author}{Mompart, J.} \& \bibinfo{author}{Corbal\'an, R.}
		\newblock \bibinfo{title}{Lasing without inversion}.
		\newblock \emph{\bibinfo{journal}{J. Opt. B: Quantum Semiclass. Opt.}}
		\textbf{\bibinfo{volume}{2}}, \bibinfo{pages}{R7--R24}
		(\bibinfo{year}{2000}).
		
		\bibitem{Kuhn2002}
		\bibinfo{author}{Kuhn, A.}, \bibinfo{author}{Hennrich, M.} \&
		\bibinfo{author}{Rempe, G.}
		\newblock \bibinfo{title}{Deterministic {Single}-{Photon} {Source} for
			{Distributed} {Quantum} {Networking}}.
		\newblock
		\href{http://dx.doi.org/10.1103/PhysRevLett.89.067901}{\emph{\bibinfo{journal}{Physical
					Review Letters}} \textbf{\bibinfo{volume}{89}}, \bibinfo{pages}{067901}}
		(\bibinfo{year}{2002}).
		
		\bibitem{Specht2011}
		\bibinfo{author}{Specht, H.~P.} \emph{et~al.}
		\newblock \bibinfo{title}{A single-atom quantum memory}.
		\newblock
		\href{http://dx.doi.org/10.1038/nature09997}{\emph{\bibinfo{journal}{Nature}}
			\textbf{\bibinfo{volume}{473}}, \bibinfo{pages}{190--193}}
		(\bibinfo{year}{2011}).
		
		\bibitem{Ritter2012}
		\bibinfo{author}{Ritter, S.} \emph{et~al.}
		\newblock \bibinfo{title}{An {Elementary} {Quantum} {Network} of {Single}
			{Atoms} in {Optical} {Cavities}}.
		\newblock
		\href{http://dx.doi.org/10.1038/nature11023}{\emph{\bibinfo{journal}{Nature}}
			\textbf{\bibinfo{volume}{484}}, \bibinfo{pages}{195--200}}
		(\bibinfo{year}{2012}).
		
		\bibitem{Reiserer2015}
		\bibinfo{author}{Reiserer, A.} \& \bibinfo{author}{Rempe, G.}
		\newblock \bibinfo{title}{Cavity-based quantum networks with single atoms and
			optical photons}.
		\newblock
		\href{http://dx.doi.org/10.1103/RevModPhys.87.1379}{\emph{\bibinfo{journal}{Reviews
					of Modern Physics}} \textbf{\bibinfo{volume}{87}},
			\bibinfo{pages}{1379--1418}} (\bibinfo{year}{2015}).
		
		\bibitem{Vitanov2017}
		\bibinfo{author}{Vitanov, N.~V.}, \bibinfo{author}{Rangelov, A.~A.},
		\bibinfo{author}{Shore, B.~W.} \& \bibinfo{author}{Bergmann, K.}
		\newblock \bibinfo{title}{Stimulated {Raman} adiabatic passage in physics,
			chemistry, and beyond}.
		\newblock
		\href{http://dx.doi.org/10.1103/RevModPhys.89.015006}{\emph{\bibinfo{journal}{Reviews
					of Modern Physics}} \textbf{\bibinfo{volume}{89}}, \bibinfo{pages}{015006}}
		(\bibinfo{year}{2017}).
		
		\bibitem{Pope2019}
		\bibinfo{author}{Pope, T.~J.} \emph{et~al.}
		\newblock \bibinfo{title}{Coherent trapping in small quantum networks}.
		\newblock
		\href{http://dx.doi.org/10.1088/1742-5468/ab54b7}{\emph{\bibinfo{journal}{Journal
					of Statistical Mechanics: Theory and Experiment}}
			\textbf{\bibinfo{volume}{2019}}, \bibinfo{pages}{124024}}
		(\bibinfo{year}{2019}).
		
		\bibitem{Villas-Boas2019}
		\bibinfo{author}{Villas-Boas, C.~J.}, \bibinfo{author}{Tolazzi, K.~N.},
		\bibinfo{author}{Wang, B.}, \bibinfo{author}{Ianzano, C.} \&
		\bibinfo{author}{Rempe, G.}
		\newblock \bibinfo{title}{Continuous Generation of Quantum Light from a Single
			Ground-State Atom in an Optical Cavity}.
		\newblock
		\href{http://dx.doi.org/10.1103/PhysRevLett.124.093603}{\emph{\bibinfo{journal}{Phys.
					Rev. Lett.}} \textbf{\bibinfo{volume}{124}}, \bibinfo{pages}{093603}}
		(\bibinfo{year}{2020}).
		
		\bibitem{Merriam2000}
		\bibinfo{author}{Merriam, A.~J.} \emph{et~al.}
		\newblock \bibinfo{title}{Efficient {Nonlinear} {Frequency} {Conversion} in an
			{All}-{Resonant} {Double}-{Lambda} {System}}.
		\newblock
		\href{http://dx.doi.org/10.1103/PhysRevLett.84.5308}{\emph{\bibinfo{journal}{Physical
					Review Letters}} \textbf{\bibinfo{volume}{84}}, \bibinfo{pages}{5308--5311}}
		(\bibinfo{year}{2000}).
		
		\bibitem{Dubin2010}
		\bibinfo{author}{Dubin, F.} \emph{et~al.}
		\newblock \bibinfo{title}{Quantum to classical transition in a single-ion
			laser}.
		\newblock
		\href{http://dx.doi.org/10.1038/nphys1627}{\emph{\bibinfo{journal}{Nature
					Physics}} \textbf{\bibinfo{volume}{6}}, \bibinfo{pages}{350--353}}
		(\bibinfo{year}{2010}).
		
		\bibitem{Mucke2010}
		\bibinfo{author}{M{\"u}cke, M.} \emph{et~al.}
		\newblock \bibinfo{title}{Electromagnetically induced transparency with single
			atoms in a cavity.}
		\newblock
		\href{http://dx.doi.org/10.1038/nature09093}{\emph{\bibinfo{journal}{Nature}}
			\textbf{\bibinfo{volume}{465}}, \bibinfo{pages}{755--8}}
		(\bibinfo{year}{2010}).
		
		\bibitem{Kampschulte2010a}
		\bibinfo{author}{Kampschulte, T.} \emph{et~al.}
		\newblock \bibinfo{title}{Optical control of the refractive index of a single
			atom}.
		\newblock
		\href{http://dx.doi.org/10.1103/PhysRevLett.105.153603}{\emph{\bibinfo{journal}{Physical
					Review Letters}} \textbf{\bibinfo{volume}{105}}, \bibinfo{pages}{153603}}
		(\bibinfo{year}{2010}).
		\newblock \bibinfo{note}{ArXiv: 1004.5348}.
		
		\bibitem{Albert2011}
		\bibinfo{author}{Albert, M.}, \bibinfo{author}{Dantan, A.} \&
		\bibinfo{author}{Drewsen, M.}
		\newblock \bibinfo{title}{Cavity electromagnetically induced transparency and
			all-optical switching using ion {Coulomb} crystals}.
		\newblock
		\href{http://dx.doi.org/10.1038/nphoton.2011.214}{\emph{\bibinfo{journal}{Nature
					Photonics}} \textbf{\bibinfo{volume}{5}}, \bibinfo{pages}{633--636}}
		(\bibinfo{year}{2011}).
		
		\bibitem{Souza2013}
		\bibinfo{author}{Souza, J.~A.}, \bibinfo{author}{Figueroa, E.},
		\bibinfo{author}{Chibani, H.}, \bibinfo{author}{Villas-Boas, C.~J.} \&
		\bibinfo{author}{Rempe, G.}
		\newblock \bibinfo{title}{Coherent {Control} of {Quantum} {Fluctuations}
			{Using} {Cavity} {Electromagnetically} {Induced} {Transparency}}.
		\newblock
		\href{http://dx.doi.org/10.1103/PhysRevLett.111.113602}{\emph{\bibinfo{journal}{Physical
					Review Letters}} \textbf{\bibinfo{volume}{111}}, \bibinfo{pages}{113602}}
		(\bibinfo{year}{2013}).
		
		\bibitem{Misra1977}
		\bibinfo{author}{Misra, B.} \& \bibinfo{author}{Sudarshan, E. C.~G.}
		\newblock \bibinfo{title}{The Zeno’s paradox in quantum theory}.
		\newblock
		\href{http://dx.doi.org/10.1063/1.523304}{\emph{\bibinfo{journal}{Journal of
					Mathematical Physics}} \textbf{\bibinfo{volume}{18}},
			\bibinfo{pages}{756--763}} (\bibinfo{year}{1977}).
		
		\bibitem{Hoffges1997}
		\bibinfo{author}{H{\"o}ffges, J.~T.}, \bibinfo{author}{Baldauf, H.~W.},
		\bibinfo{author}{Eichler, T.}, \bibinfo{author}{Helmfrid, S.~R.} \&
		\bibinfo{author}{Walther, H.}
		\newblock \bibinfo{title}{Heterodyne measurement of the fluorescent radiation
			of a single trapped ion}.
		\newblock
		\href{http://dx.doi.org/10.1016/S0030-4018(96)00621-9}{\emph{\bibinfo{journal}{Optics
					Communications}} \textbf{\bibinfo{volume}{133}}, \bibinfo{pages}{170--174}}
		(\bibinfo{year}{1997}).
		
		\bibitem{Laflamme2017}
		\bibinfo{author}{Laflamme, C.}, \bibinfo{author}{Yang, D.} \&
		\bibinfo{author}{Zoller, P.}
		\newblock \bibinfo{title}{Continuous measurement of an atomic current}.
		\newblock
		\href{http://dx.doi.org/10.1103/PhysRevA.95.043843}{\emph{\bibinfo{journal}{Physical
					Review A}} \textbf{\bibinfo{volume}{95}}, \bibinfo{pages}{043843}}
		(\bibinfo{year}{2017}).
		
		\bibitem{Harbola2012}
		\bibinfo{author}{Harbola, U.}, \bibinfo{author}{Rahav, S.} \&
		\bibinfo{author}{Mukamel, S.}
		\newblock \bibinfo{title}{Quantum heat engines: {A} thermodynamic analysis of
			power and efficiency}.
		\newblock
		\href{http://dx.doi.org/10.1209/0295-5075/99/50005}{\emph{\bibinfo{journal}{EPL
					(Europhysics Letters)}} \textbf{\bibinfo{volume}{99}},
			\bibinfo{pages}{50005}} (\bibinfo{year}{2012}).
		
		\bibitem{HamsenThesis}
		\bibinfo{author}{Hamsen, C.}
		\newblock \emph{\bibinfo{title}{Interacting Photons in a Strongly Coupled
				Atom-Cavity System}}.
		\newblock \bibinfo{type}{Dissertation}, \bibinfo{school}{Technische
			Universität München}, \bibinfo{address}{München} (\bibinfo{year}{2017}).
		
	\end{thebibliography}

\end{document}


\title{Supplementary Information: \\Continuous quantum light from a dark atom}
	
	\author{Karl Nicolas Tolazzi}
	\email{nicolas.tolazzi@mpq.mpg.de}
	\author{Bo Wang}
	\author{Christopher Ianzano}
	\author{Jonas Neumeier}
	\affiliation{Max-Planck-Institut f\"ur Quantenoptik, Hans-Kopfermann-Str.\ 1, D-85748 Garching, Germany}
	\author{Celso Jorge Villas-Boas}
	\affiliation{Departamento de F\'isica, Universidade Federal de S\~ao Carlos, P.O. Box 676, 13565-905, S\~ao Carlos, Brazil}
	\author{Gerhard Rempe}
	\affiliation{Max-Planck-Institut f\"ur Quantenoptik, Hans-Kopfermann-Str.\ 1, D-85748 Garching, Germany}
	
	\date{\today}
	\maketitle
	\newpage
	
	\section{Simulations}
	In this section, the simulations and the concrete implementations from the main text are described. We present the Hamiltonian of the ideal system and how we simulate correlation functions.
	
	\subsection{Hamiltonian and photon correlations}
	Here, we choose $\hbar = 1$ for simplicity. The general Hamiltonian is then given by:
	\begin{align}\label{eqn:Hamilt}
		\hat{H} = \hat{H}_{\text{atom}} + \hat{H}_{\text{cavity}} + \hat{H}_{\text{int}} + \hat{H}_{\text{d}}.
	\end{align}
	$\hat{H}_{\text{atom}}$ and $\hat{H}_{\text{cavity}}$ are the Hamiltonians for the unperturbed atom and cavity. The interaction between an atomic transition and a cavity mode is described by $\hat{H}_{\text{int}}$. 
	
	For a three-level atom and a cavity mode we find:
	\begin{align}
		\hat{H}_{\text{atom}} &= \omega_2 \hat{\sigma}_{22} + 
		\omega_3 \hat{\sigma}_{33},  \\
		\hat{H}_{\text{cavity}} &= \omega_{\text{cav}} \hat{a}^\dagger \hat{a}.
	\end{align}
	Here the zero point of energy is chosen at the level of state $\ket{1}$. With the dipole and rotating wave approximations the interaction Hamiltonian is given by:
	\begin{align}
		\hat{H}_{\text{int}} &=  g \left( \hat{a}^\dagger \hat{\sigma}_{13} +  \hat{\sigma}_{13}^\dagger \hat{a}\right).
	\end{align}
	
	The driving term $\hat{H}_{\text{d}}$ is given by:
	\begin{align}
		\begin{split}
			\hat{H}_{\text{d}} &=\frac{\Omega_{12}}{2} e^{-i \omega_{12}t} \hat{\sigma}_{12}\\
			&+\frac{\Omega_{23}}{2} e^{-i \omega_{23}t} \hat{\sigma}_{23}
			+ h.c..
		\end{split}
	\end{align}
	$\Omega_{12,23}$ are the Rabi frequencies. $\omega_{12,23}$ are the frequencies of the corresponding laser beams.

	In order to simulate the photon correlations which are measured in the experiment, dissipative processes need to be included in the simulation. This is done by solving the Lindblad master equation:
	\begin{align}\label{eqn:master}
		\dot{\rho(t)} =& -\frac{i}{\hbar}\left[\hat{H},\rho(t)\right] \nonumber \\
		&+\sum_{i} \left(2\hat{C}_i\rho(t)\hat{C}^\dagger_i-\rho(t)\hat{C}^\dagger_i\hat{C}_i - \hat{C}^\dagger_i\hat{C}_i\rho(t)\right).
	\end{align}
	Here, $\rho(t)$ is the density matrix and $\hat C_i$ are the dissipation operators that can be categorized into atomic decays
	\begin{subequations}
		\begin{align}
			\hat C_{1} &= \sqrt{\gamma_{13}}\hat \sigma_{13},\\
			\hat C_{2} &= \sqrt{\gamma_{23}}\hat \sigma_{23},
		\end{align}
		with the decay constants $\gamma_{i3}$ from the atomic state $\left|3\right\rangle$ to the atomic state $\left| i \right\rangle$, cavity decay
		\begin{align}
			\hat C_{3} = \sqrt{\kappa} \hat a,
			\label{cavitydecay}
		\end{align}
		with decay constant $\kappa$, and the dephasing between the two ground states
		\begin{align}
			\hat C_{4} = \sqrt{\gamma_d} \hat{\sigma}_{22},
			\label{dephasing}
		\end{align}
		with dephasing rate $\gamma_d$.
	\end{subequations}
	We can rewrite Equation \ref{eqn:master} using the Lindblad super-operator $\mathcal{L}$:
	\begin{align}
		\dot \rho = \mathcal{L} \rho
	\end{align}
	and formally solve it with:
	\begin{align}
		\rho\left( t \right) = e^{\mathcal{L}t}\rho\left( 0 \right).
	\end{align}
	The density matrix of the steady state $\rho_{\text{ss}}$ can be found by imposing:
	\begin{align}
		\mathcal{L} \rho_{\text{ss}} = 0.
	\end{align}
	
	The expectation values for an operator $\hat O$ are then given by:
	\begin{align}
		\mean{\hat O(t)} = \text{Tr}\left( \hat O e^{\mathcal{L}t} \rho(0) \right).
	\end{align}
	According to quantum regression theorem, the photon correlation can now be determined by: 
	\begin{align}
		\mean{\hat a^\dagger \hat a^\dagger\left(\tau \right) \hat a \left(\tau \right) \hat a}
		=\text{Tr}\left(\hat a^\dagger \hat a e^{\mathcal{L}\tau} \left( \hat a \rho_{ss} \hat a^\dagger \right)\right).
	\end{align}

	\subsection{Zeno blockade}
	The ladder of dark states exhibits a Zeno blockade that is more or less pronounced depending on $\Omega_{23}$. To see this, we determine the decay rate, which can be derived via Fermi's golden rule to: 
	\begin{align}
		\begin{split}
			\Gamma_{n \rightarrow n-1} &=\left|\left\langle \Psi_{n-1}^{0}\right|\sqrt{\kappa}a\left\vert \Psi_{n}^{0}\right\rangle \right|^{2} \\
			&=\frac{\kappa n\left[4 g^{2}\left(n-1\right)+\Omega_{23}^{2}\right]}{\left(4 g^{2}n+\Omega_{23}^{2}\right)}.
		\end{split}
	\end{align}
	We must then compare this to the effective driving strength for the transition  $n-1 \rightarrow n$, that is given by:
	\begin{align}
		\Omega_n=\frac{2 \Omega_{12}\Omega_{23}g\sqrt{n}}{\sqrt{4 g^{2}n+\Omega_{23}^{2}}\sqrt{4g^2(n-1)+\Omega_{23}^2}}.
	\end{align}
	The derivation of both formulas can be found in reference \cite{Villas-Boas2019}.
	We can now define a Zeno-factor, the driving strength in comparison to the decay rate, as:
	\begin{align}
		Z_n=\Omega_n/\Gamma_{n}.
	\end{align}
	
	\begin{figure}[t]
		\includegraphics[width=8.6cm]{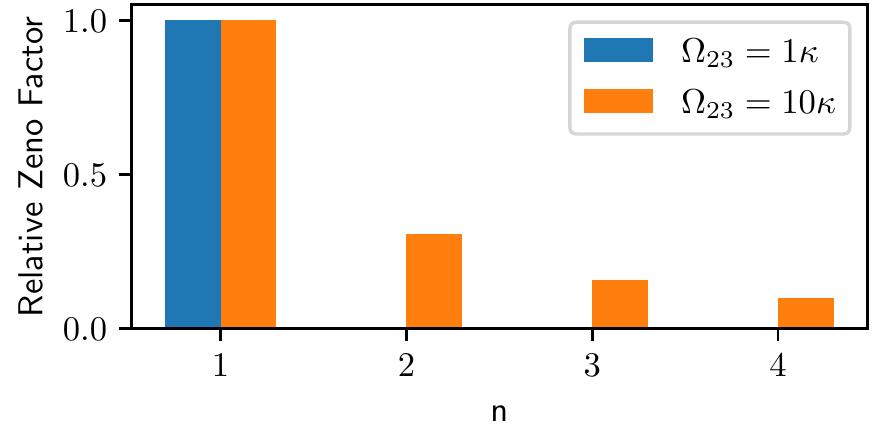}
		\caption{\label{fig_zeno}The relative Zeno factor as defined in the text for different rungs $n$ and two different values of $\Omega_{23}$.}
	\end{figure}
	
	Fig. \ref{fig_zeno} shows now the relative Zeno factor (normalized to Z(1)) for two different values of $\Omega_{23}$. For a low value of $\Omega_{23}=\SI{1}{\mega\hertz}$, the blockade of higher rungs of the dark state ladder becomes clear. $Z_n$ for $n>1$ is so small that it is not even visible in the plot and $Z_2/Z_1$ is around $10^{-3}$. Thus the system really behaves like a two level system, showing only dynamics between  $\left\vert \Psi_{0}^{0}\right\rangle$ and $\left\vert \Psi_{1}^{0}\right\rangle$. If $\Omega_{23}\sim g$, there is still suppression but no blockade anymore as can be seen in \ref{fig_zeno}. This shows that we do indeed have a tunable Zeno blockade where it is possible to either cut the Hilbert space after the first excitation, or to climb the dark state ladder. 
	
	\subsection{Excitation of the atom}
	The main paper states that the system produces photons from dark states while avoiding the atomic excited state. Therefore we designate as the figure of merit $\langle a^\dagger a \rangle/\langle \sigma_{33} \rangle$. This factor gives the average cavity population per average excitation in the atomic excited state $\ket{3}$. 
	\begin{figure}[ht]
		\includegraphics[width=8.6cm]{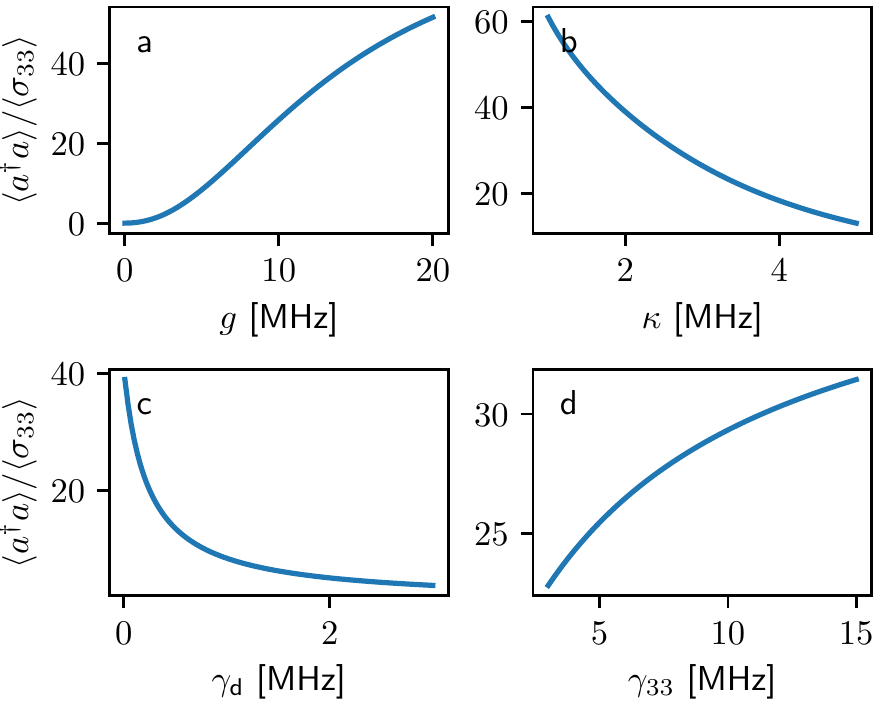}
		\caption{
			\label{fig_excitationslessness}
			$\langle a^\dagger a \rangle/\langle \sigma_{33} \rangle$, plotted against different variables of the system. All other parameters are fixed to the experimental values of $g/2\pi=\SI{9.2}{\mega\hertz}, \Omega_{12}/2\pi=\SI{0.3}{\mega\hertz}, \Omega_{23}/2\pi=\SI{4.0}{\mega\hertz}, \gamma_d/2\pi=\SI{0.13}{\mega\hertz}$.}
	\end{figure}
	Fig. \ref{fig_excitationslessness} shows this factor against different variables of the system. Worth noting is that this figure of merit increases with $g$, but also increases with the excited state decay rate $\gamma_{33} = \gamma_{31}+\gamma_{32}$ for certain parameters. Furthermore the factor decreases with a higher $\kappa$ and higher dephasing rate between the ground states $\gamma_{d}$.

	\subsection{The real atom and uncoupled states}
	Whenever there is population in the excited state of the atom it becomes possible that it decays to states that are uncoupled by any laser field. This means there is a total number of photons that the system can scatter before it is fully decayed into one of these uncoupled states. This number scales with the figure of merit, the photon generation per excited state population $\langle a^\dagger a \rangle/\langle \sigma_{33}\rangle$. If there would be no population in the excited state at all, the amount of photons produced would conceivably be infinite.
	
\begin{figure}[ht]
		\includegraphics[width=8.6cm]{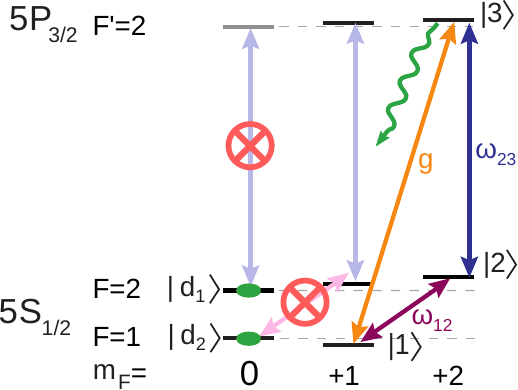}
		\caption{
			\label {fig_6level}
			The three-level system, $\left\{\ket{1}, \ket{2}, \ket{3}\right\}$, is implemented on Zeeman substates of the $D_2$ Line of the $^{87}\text{Rb}$ atom. The cavity coupled transition is indicated by an orange arrow. Once the other Zeeman sublevels are considered, the transition becomes open, allowing the atom to decay into an uncoupled state. The atomic population will eventually accumulate in the states $\ket{\text{d}_1}$ and $\ket{\text{d}_2}$.}
	\end{figure}

\begin{figure}[th]
		\includegraphics[width=8.6cm]{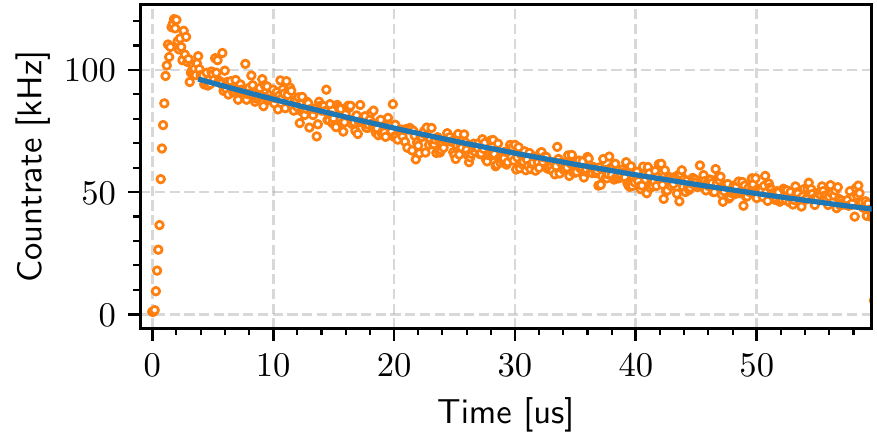}
		\caption{
			\label {fig_depumping}
			Countrate of the system at $\Delta_{12}=\Delta_{23}=0$ (as defined in the main text) over time after state preparation and starting the experiment at time 0. The system decays into one of the uncoupled states after some time because of residual excitation in state $\ket{3}$.}
	\end{figure}

	For further discussion it is necessary to look at the concrete implementation in a real atom.
	The general level scheme is implemented on the $D_2$ line of the $^{87}\text{Rb}$ atom. The states $\ket{1},\ket{2},\ket{3}$ are mapped to real atomic states as follows:
	\begin{align*}
		\ket{1}&=\ket{5\text{S}_{1/2},\text{F}=1,\text{m}_\text{F}=+1},\\
		\ket{2}&=\ket{5\text{S}_{1/2},\text{F}=2,\text{m}_\text{F}=+2},\\
		\ket{3}&=\ket{5\text{P}_{3/2},\text{F}'=2,\text{m}_\text{F'}=+2}.
	\end{align*}
	The aforementioned theory that considered three atomic levels is not sufficient to explain why the measured count rate in the system decreases over time. The real $^{87}$Rb atom is much more complex and the excited state can decay into different ground states as can be seen in Fig. \ref{fig_6level}. We start with a system that is well prepared in ground state $\ket{1}$ but after excitation to state $\ket{3}$ the system can also decay out of the cycle into the state $\ket{5\text{S}_{1/2},\text{F}=2,\text{m}_{\text{F}}=1}$.
	From there it can be pumped up again to the state $\ket{5\text{P}_{3/2},\text{F}'=2,\text{m}_{\text{F'}}=1}$ by the (then slightly detuned) laser with $\omega_{23}$. It can then either fall back into the original cycle via the $\ket{5S_{1/2},\text{F}=2,\text{m}_\text{F}=2}$ state (without producing photons in the cavity) or decay terminally to state $\ket{\text{d}_1}=\ket{5\text{S}_{1/2},\text{F}=2,\text{m}_{\text{F}}=0}$ or state  $\ket{\text{d}_2}=\ket{5\text{S}_{1/2},\text{F}=1,\text{m}_{\text{F}}=0}$. These are uncoupled states to the system because they are dipole forbidden in case $\ket{\text{d}_1}$, for the only resonant laser $\omega_{23}$, or are detuned due to Zeeman-splitting in case $\ket{\text{d}_2}$. The atomic population will be accumulated in these states and the system won't scatter further photons until reinitialized. 
	This leads to a decay of the measured photon production rate over time as shown in Fig. \ref{fig_depumping}. This specific measurement produced 4.0 photons in total. As is clearly visible in Fig. \ref{fig_depumping}, the depumping process is not completed in the measurement duration of \SI{60}{\micro\second} and even more photons could potentially be produced for a longer measurement interval. This maximum number of photons that the system can produce before falling into an uncoupled state is the number of relevance. This number can be extrapolated easily from the exponential decay of the countrate to 6.9(1) photons for this measurement.
	
	In order to simulate this system, both with and without a cavity, a full quantum Monte-Carlo simulation with all the 7 relevant levels and appropriate decay channels, reentry channels and uncoupled states was implemented. This allowed not only for the investigation of the theoretical number of photons that can be scattered by our system, but for the number that would be possible to scatter in such a system in the absence of a cavity. The number of expected photons in an equivalent atomic system that lacks a cavity (or in the case of the Monte-Carlo simulation, with the cavity coupling constant, $g$, set to 0) is shown to be 1.55. For both simulations, only photons with cavity transition frequency were considered ($\ket{\text{F}=1}\rightarrow \ket{\text{F}'=2}$). The cavity-less simulation assumes a 4$\pi$ solid angle measurement, and a detection efficiency equivalent to our experimental one. This detection efficiency is primarily the product of a 57\% cavity outcoupling efficiency, a 65\% detector efficiency, and an 80\% fiber incoupling efficiency for the capture of the cavity mode, with approximately 90\% efficiency from the remaining propagation. Our net efficiency is then approximately 26\%.

	\subsection{Limiting factors}
	This section briefly describes which are the limiting factors to the photon enhancement as discussed in the previous chapter and how the enhancement factor could be improved.
	As is explained in the main text, we trap a single Rubidium atom in a three-dimensional optical lattice in the focus of a high-finesse optical resonator. We load this atom via an atomic fountain from below, which leads to a probabilistic distribution of the atomic position within the cavity. The trapping time for each atom is only of the order of 6s, so  each presented measurement is a statistical mixture of different atomic positions. Different positions are accompanied by variations of many parameters. First of all, the coupling constant g is varying to a small extend due to different overlaps between trapping sites and the cavity field. Furthermore the driving strengths of all the lasers vary depending on the position, as the lasers have a fairly small beam waist to be able to pass through the cavity from the side without being clipped and to provide sufficient power density, and therefore Rabi frequency, at the atom’s position. These problems are only technical in nature and could be mitigated by implementation of a deterministic loading scheme like loading via transport in a conveyor belt. 
	Furthermore, multiple other factors limit the total amount of photons. First of all, there is imperfect state preparation to the initial state, which is notoriously hard to prepare. At the moment, this is based on incoherent optical pumping, which only brings around 80\% of the population to the desired state. This means that approximately 20\% of the time, we don’t detect any photons in the cycle because the initial state is incorrect. This can be improved by using coherent state preparation via e.g. a pi-pulse from a state that can be prepared with a better fidelity. This could bring almost 100\% of the population into the desired state as other experiments have shown. 
	A second major limitation for the enhancement factor of photons can already be seen in \autoref{fig_excitationslessness}(c). The two ground states in the cycle experimentally exhibit a dephasing rate $\gamma_d$ due to magnetic field fluctuations. This is deadly for the dark states as they are based on coherent ground state superpositions. This leads to fewer photons in total, as the dark state becomes “less dark”. This could be mitigated by improving the magnetic field shielding which has led to a major improvement in many other sensitive quantum experiments as well. 
	Finally, also the coupling of the two ground states via Raman transitions is not optimal, as the necessary high power of the beams leads to a substantial scattering rate of as high as 10kHz (from a theoretical calculation) despite the detuning from any atomic transition.  This results in decoherence and even depumping into uncoupled states. This could be improved by using a direct microwave transition between the two states. We did not implement this in our setup due to limited access for a microwave field but this is, in principal, a technical limitation, not a physical one.

	\bibliographystyle{naturemag}